\documentclass{article}

\usepackage{amsmath}
\usepackage{amssymb,amsfonts,textcomp}
\usepackage{xcolor}
\usepackage{adjustbox}
\usepackage{caption}
\usepackage{subcaption}
\usepackage{array}
\usepackage[numbers]{natbib}
\usepackage{float}
\usepackage{xurl}
\usepackage{ragged2e}
\usepackage[raggedright]{titlesec}

\newcommand{\referenceSection}{
	\section*{REFERENCES}
	\renewcommand{\section}[2]{}%
	\def\bibindent{1em}
	\def\labindent{1em}
}

\title{A Characterization of 3D Printability}

\begin{document}

\maketitle

\centering{
	\author{Ioannis Fudos}\textsuperscript{1},
	\author{Margarita Ntousia}\textsuperscript{1}, 
	\author{Vasiliki Stamati}\textsuperscript{1}, 
	\author{Paschalis Charalampous}\textsuperscript{2},
	\author{Theodora Kontodina}\textsuperscript{2},
	\author{Ioannis Kostavelis}\textsuperscript{2},
	\author{Dimitrios Tzovaras}\textsuperscript{2},
	\author{Leonardo Bilalis}\textsuperscript{3}
}

\bigskip

\centering{
\textsuperscript{1}{ Dept. of Computer Science \& Engineering, University of Ioannina}{\{fudos, mntousia, vstamati\}@cse.uoi.gr},
\textsuperscript{2}{Centre for Research and Technology Hellas, Information Technologies Institute}{\{pcharalampous, kontodinazoli, gkostave, Dimitrios.Tzovaras\}@iti.gr},
\textsuperscript{3}{3D Life}{leonardo.bilalis@3dlife.gr}
}


\begin{abstract}
 
 Additive manufacturing technologies are positioned to provide an unprecedented innovative transformation in how products are designed and manufactured. Due to differences in the technical specifications of AM technologies, the final fabricated parts can vary significantly from the original CAD models, therefore raising issues regarding accuracy, surface finish, robustness, mechanical properties, functional and geometrical constraints. Various researchers have studied the correlation between AM technologies and design rules.

In this work we propose a novel approach to assessing the capability of a 3D model to be printed successfully (a.k.a \textit{printability}) on a specific AM machine. This is utilized by taking into consideration the model mesh complexity and certain part characteristics. A $printability$ $score$ is derived for a model in reference to a specific 3D printing technology, expressing the probability of obtaining a robust and accurate end result for 3D printing on a specific AM machine. The printability score can be used either to determine which 3D technology is more suitable for manufacturing a specific model or as a guide to redesign the model to ensure printability. We verify this framework by conducting 3D printing experiments for benchmark models which are printed on three AM machines employing different technologies: Fused Deposition Modeling (FDM), Binder Jetting (3DP), and Material Jetting (Polyjet). 
\end{abstract}

\noindent
\justify


\section{INTRODUCTION}
Additive Manufacturing (AM) is currently being promoted as the spark of a new industrial revolution, due to its versatility in creating 3D structures of unprecedented design freedom and geometric complexity in comparison with conventional manufacturing techniques~\cite{Tofail2018,Gibson}. AM refers to a great variety of commercially available technologies that are most widely applied to manufacture 3D models directly from CAD data, based on successive layer deposition of material in a pre\text{-}arranged pattern. 
Due to differences in AM technologies, in regards to employed processes, machines and materials, the final fabricated part can vary, sometimes significantly, from the originally designed one, therefore raising issues regarding dimensional accuracy, surface finish, mechanical properties, functional and geometrical requirements~\cite{Kim2018, Oropallo2016}. Various research efforts have articulated the correlation between AM technologies and the design process, in terms of integrating specific design rules or guidelines pertinent to model complexity, design potentials and constraints of each AM process~\cite{Yang2015, Tompson2016}. 

In this work we study, determine and correlate the complexity and the part characteristics of a CAD model with its ability to be printed - a.k.a. {\em printability} - using a specific printing technology. This is accomplished mainly in terms of structural robustness and dimensional accuracy of the corresponding 3D model. We propose a novel approach that computes a $printability$ $score$  for a specific 3D printing technology by taking into consideration the model mesh complexity and certain part characteristics. This score expresses the probability of producing a robust and accurate end result on a specific AM machine. To achieve this we isolate part design rules often used in design guidelines for AM, whose parameter values directly affect the quality of a manufactured model and map them to probability functions that, when evaluated, produce a printability score. This metric can be used either to determine which 3D technology is more suitable for manufacturing a specific model or as a guide to redesign the model to ensure printability. We verify this measure by conducting printing experiments for several benchmark models as test cases on three AM machines employing different technologies: Fused Deposition Modeling (FDM), Binder Jetting (3DP), and Material Jetting (Polyjet). 

The proposed framework is not restricted to these 3D printing technologies and can be parameterized and/or extended to estimate printability for other AM technologies. 

The rest of this paper is structured as follows. Section 2 provides a brief survey of related work and Section 3  offers an overview of model and part characteristics that may affect the printability of a model. Section 4 introduces an innovative measure that characterizes printability through a novel mathematical formulation that estimates the probability that a model will be printed correctly. Section 5 provides an experimental validation of our printability measure. Finally, Section 6 offers conclusions.

\section{RELATED WORK}
\label{sec:relatedwork}
3D models, as detailed representations of solids or surfaces, are commonly used in several domains such as medicine, engineering, analysis, manufacturing, arts etc \cite{stavropoulos20103}. One critical 'parameter' that affects several functions and processes of these models, such as design, engineering and manufacturing, is the complexity of the model. Despite the fact that there is no standard objective complexity metric, several researchers over the years have followed different approaches to define and measure model complexity mainly for purposes of implementation time and cost reduction. It is also widely accepted that the more complex the model is, the less robust and flexible it is. Also, a CAD model with high complexity makes it difficult for adjustments, modifications and analysis, but also affects production time and cost. That is why it is important to reduce the complexity of a model by suppressing the less significant features and consequently reducing the data describing these features in the CAD model~\cite{Rossignac2005,Globa2016,Johnson2018}.

Even though geometrical and CAD model complexities might be related, they are not the same, since the second one refers to the complexity of the CAD model which is used to represent one component, its features and the relationships between them. Over the years several researchers studied model classification, with Forrest~\cite{Globa2016} suggesting 3 types of complexity starting with geometrical complexity, referring to basic elements such as points, lines, surfaces, etc., combinatorial complexity, which counts the elements of a model, the number of vertices in a polynomial mesh, edges, faces, etc., and dimensional complexity, which characterizes a model as 2D, 2.5D or 3D.

Other complexity metrics discussed over the years are algebraic, topological, morphological and combinatorial
\cite{Rossignac2005}.
Algebraic complexity deals with the complexity degree of the polynomials required to represent the exact shape of a model. In  Constructive Solid Geometry (CSG) a complex solid or surface object can be created by combining simpler primitive objects such as boxes, cylinders, spheres, pyramids etc. through Boolean operations. Surfaces of CSG primitives are also defined by polynomials. In the case of free form/sculptured shapes, a larger amount of quadric primitives is required for a closer approximation since the CSG model is not enough. CSG can also be performed on polygonal meshes, and may or may not be procedural and/or parametric. 
Topological complexity deals with 3D geometries, such as surfaces, that are not boundaries of a solid, models with internal structure, non-regularized shapes, measures the existence of holes, non-manifold singularities, self-intersections and finally the genus complexity.
Morphological complexity deals with the number of features of a shape, its size, smoothness and regularity. Triangular mesh subdivision can increase its smoothness. Through the subdivision process, high-resolution models are achieved; the number of vertices on the model is increased, leading to more rounded curves.
Combinatorial complexity also deals with reducing the number of vertices in a triangular mesh. By reducing the combinatorial complexity the model might be simpler but also less accurate.

Another approach uses the number of surfaces a component has as complexity metrics; the larger the number, the more complex the model is. The number of triangles in the STL file used for component representation is another assumption of geometric complexity leading to high quality fabricated parts. Other approaches calculate geometric complexity by comparing the volume of the component with the volume of its bounding box. Both the number and the shape of the features forming a model affect the geometric complexity, since a large number of irregular, thin or even curved features increases not only the computed printing time but also the final form of the fabricated part~\cite{Johnson2018}.

The quality of a fabricated part is directly linked to the design principles and rules followed during the design phase, prior to printing. There is a plethora of work that evaluate AM processes and correlates them with design for additive manufacturing (dfAM)\cite{Booth2016}. Design rules have been defined for specific 3D printing technologies, to help designers produced parts of high quality and in conformance with the initial design. In \cite{Mani2017} a review is provided on design principles that have been defined based on design rules for AM and a Guide\text{-}to\text{-}Principle\text{-}to\text{-}Rule (GPR) approach is proposed to assist the design process for better manufacturing. Design rules are grouped together in \cite{Jee2015DESIGNRW} to form modules as a more dynamic and designer-friendly way of dealing with the design process and a case study is presented on powder bed fusion technology. Design rules for AM were developed in \cite{Adam2015} based on geometrical standards and attributes that characterize the object shapes. Three different technologies were used to manufacture the elements for different attribute values and a design catalogue is created. In \cite{Zhang2016} an analysis  of design features in AM is provided and a feature-based approach to orientating parts for optimized building in AM is presented.

The subject of print failure is addressed in ~\cite{Booth2016}. The authors designed and implemented a dfAM worksheet that is used either in the conceptual phase or the CAD phase of the design process. The use of the worksheet led to a decrease in print failures. 

Our work focuses on evaluating the printability of a model on a specific 3D printing technology, following the design phase. It does not use design rules to guide the design process, as is more often the case in the aforementioned literature, but uses them as an evaluation guideline to help predict the quality of the 3D print on a certain AM machine and support the decision making. It could, however, be used as knowledge for re-designing a model to improve the printability on a specific AM machine.

For the purposes of the present work and in order to measure the printability of a model the geometrical complexity will be taken into account regarding the number of triangles in the STL file which affect the regularity and smoothness of the final surface. The volumes of the different triangulated models and the corresponding Bounding Boxes will be compared with the initial CAD model from which the deviation rate will be derived. As for the geometric features of the model, such as wall thicknesses, self supporting angles and holes\cite{ComputerAided}, they play a significant role and will be taken into account for the evaluation of printability. 



\section{MODEL AND PART CHARACTERISTICS THAT AFFECT PRINTABILITY}
\label{sec:characteristics}

The quality of a 3D printed model, in reference to its robustness and conformance to the initial model, depends on various parameters. In the present work we focus on the 3D representation of the model which is used for printing and specific design characteristics of the model that might affect its printability. 

\subsection{Geometric Primitives and Benchmarks}
To evaluate both the design and the printing processes we used models of geometric primitives and benchmark models designed especially for this purpose. The geometric primitives included spheres, cylinders, tori and rectangular parallelepipeds, and triangular meshes of different mesh resolutions were created for each primitive. 
Also, three different benchmarks were designed to test and evaluate different design characteristics, printing capabilities and limits. As shown in Table ~\ref{tbl:threeBenchmarks}, the first benchmark (B1) contains supported and unsupported walls and horizontal bridges of different thicknesses (THK), through holes of various diameters (dia), embossed and engraved details of different widths (w) and heights (h), as well as many other details. This benchmark was constructed based on specific design rules for different 3D printing technologies, as they have been recorded in \cite{3DHubsDesign}, and examines marginal values of basic geometrical features, surface roughness and dimensional accuracy. The second benchmark (B2) also contains supported and unsupported walls of different THK, overhangs of varying angles and several connecting parts such as a sphere, propeller, cover and more, as shown in detail in Figure ~\ref{fig:Bench-prints}(b), later on in Section \ref{sec:validation}. The construction was based on reports from 10 years of experience in research and implementation of failed designs of 3D Life company \cite{3DLife}, once more to examine marginal values of basic geometrical features, test surface roughness and dimensional accuracy, functionality  and precise connectivity between the different detachable parts. Finally, the third benchmark (B3) is constructed using basic CSG operations and contains basic geometrical features from various models \cite{Decker2015}, \cite{Rebaioli2017}, \cite{Minetola2016} supported walls of, embossed and engraved details of different w and h, to test, once more, surface roughness and dimensional accuracy.

\begin{table}[h!]
\centering
\begin{adjustbox}{width=1\textwidth}
\begin{tabular}{c|c|c} \hline
  B1 benchmark & B2 benchmark & B3 benchmark \\ \hline
  \captionsetup{justification=centering}
  \includegraphics[width=0.3\textwidth]{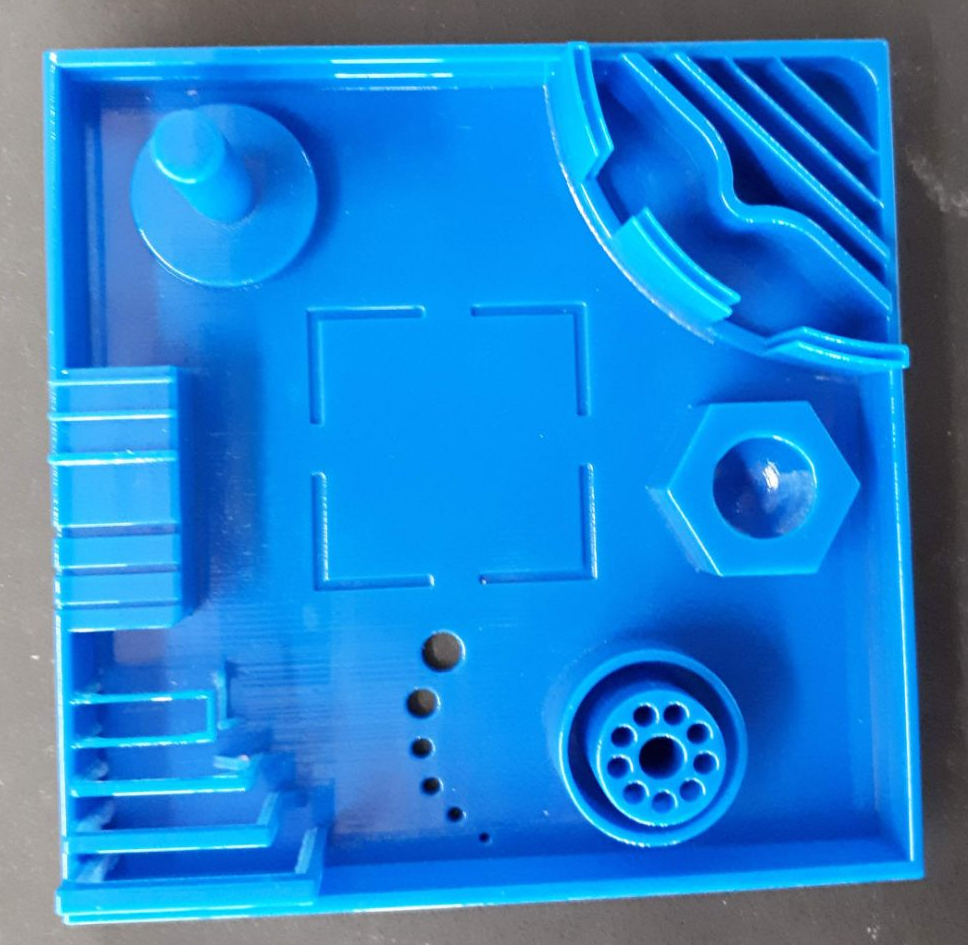}  &  \captionsetup{justification=centering}
  \includegraphics[width=0.4\textwidth]{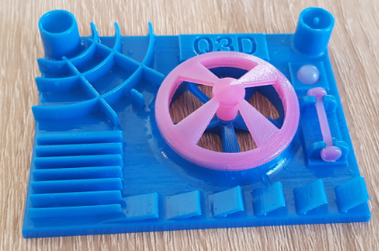}  & \captionsetup{justification=centering}
  \includegraphics[width=0.3\textwidth]{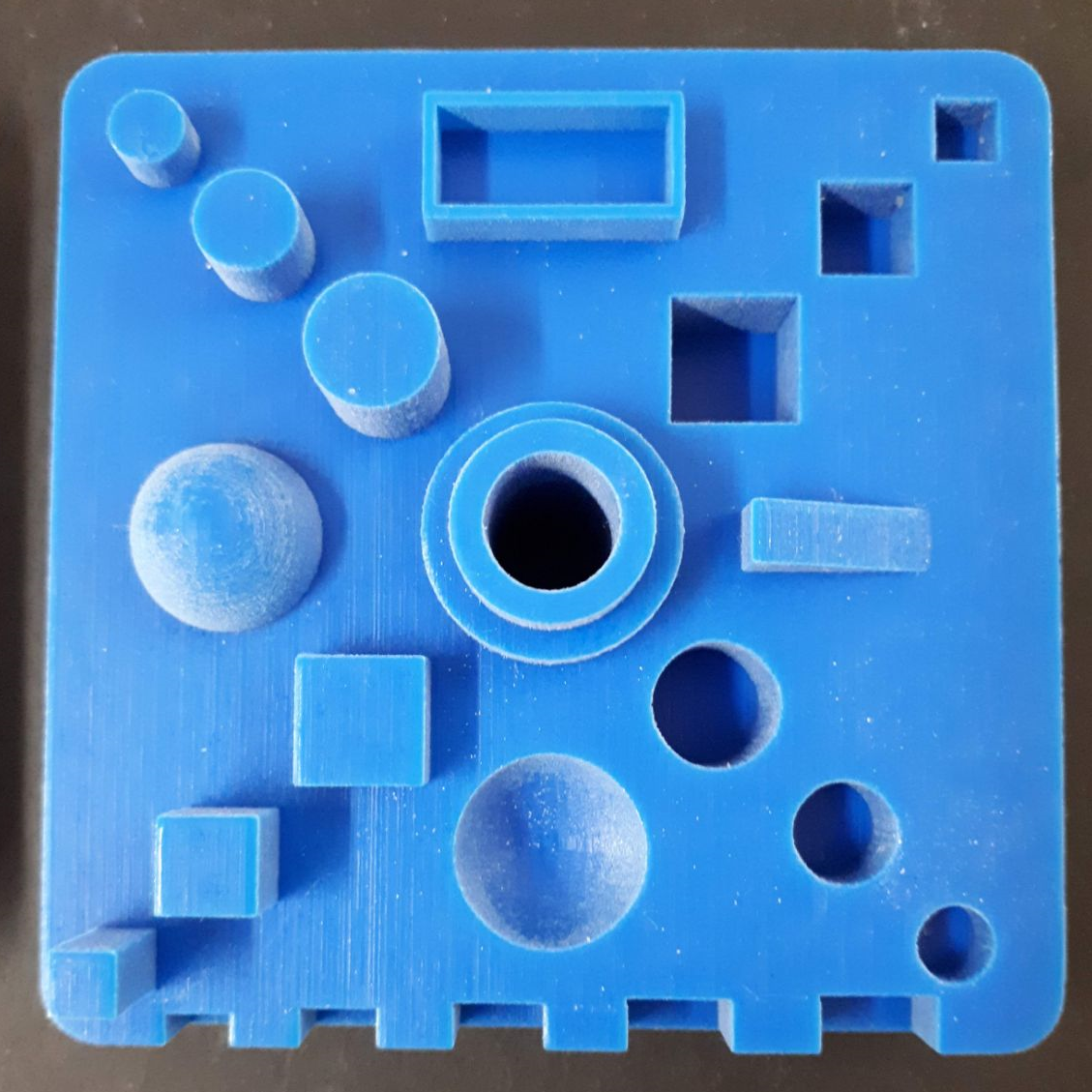} \\ \hline
Supported walls: 0.3 \textless THK \textless 1.5:mm & Unsupported walls: 0.3 \textless THK \textless 2.0:mm & Supported walls: THK = 1.0:mm\\ 
Unsupported circular wall:  THK = 1.0:mm & Unsupported walls: 10$^{\circ}$ \textless angles \textless 50$^{\circ}$ &\\ 
Through holes: 0.2 \textless dia \textless 2.0:mm & Supported walls: THK = 1.0:mm &\\ 
Horizontal bridges: 0.3 \textless THK \textless 2.0:mm & Engraved details: h, w = 1.0:mm & \\ 
Embossed/Engraved details: 0.5 \textless h, w \textless 1.0:mm & Pin dia: 1.0:mm\\ 

\end{tabular} 
\end{adjustbox}
\caption{Benchmark's design rules and restrictions.}
\label{tbl:threeBenchmarks}
\end{table}

\subsection{Mesh Complexity}
In order to measure the printability of the designed CAD models, the geometrical complexity will be measured as the complexity of the mesh representation of the CAD model which is submitted to the AM machine for slicing (and/or {\em g-code} production) and printing, and therefore referred to as \textit{mesh complexity}. After the design process is complete, the CAD models are exported as polygon meshes, commonly in the form of triangles, and stored in an STL file. We define the mesh complexity $C$ of a CAD model $M$ that is converted to a mesh that consists of a set of triangles {\textit{PLG(M)}} that approximates the initial CAD model as:
\begin{equation}
\label{eq:mesh1}
C_{M} = |PLG(M)|
\end{equation}

If the mesh consists of convex polygons (not necessarily triangles), then each convex polygon $p$ with $v(p)$ vertices is triangulated and the overall mesh complexity is:
\begin{equation}
\label{eq:mesh2}
C_{M} = \sum_{p \in M}{v(p)-2}
\end{equation}

For each of the CAD geometric primitives and benchmarks, four triangulations of different resolutions were calculated, starting from a low quality representation to a highest quality almost matching the initial geometry of the models. For each of these triangulations the new volumes and bounding boxes were calculated and the corresponding graphs were plotted to display the deviation between the meshed and the initial CAD model. A Mean Curvature Analysis was performed on all the initial and triangulated models to measure the average value between Minimum and Maximum curvatures and the values of the different curvatures for each vertex are summarized in histograms for a visual representation of the data distribution. 

As an example, the models of the sphere and B3 benchmark are presented in more detail. The various measurements of the meshed models are illustrated in Tables~\ref{tab:spheregeo} and \ref{tab:B3geo}. The resolution of a mesh is directly correlated to the quality, level of detail and robustness of its corresponding printed form. Models approximated by lower resolution meshes produce printed models that exhibit significant deviations from the original model through loss of detail and surface quality reduction. Mesh complexity is also related to the morphology of the model. Surfaces of high curvature must be represented by higher resolution meshes to better approximate the initial CAD model. For instance, the triangulation of a CAD model leading to a representation with 168 faces causes a reduction in the model volume over 8\%, which decreases as the number of elements increases. As such, for a model with 148224 faces the percentage of loss asymptotically reaches 0\%.

\begin{table}[h]
\centering
\begin{tabular}{c|c|ccc} \hline 
  \multicolumn{5}{c}{Sphere $D\colon 3$ cm - Volume$\colon 1.414*10^{-5}$ $m^3$}  \\ \hline
  Triangular Mesh & Volume ($10^{-5}$ $m^3$) &  \multicolumn{3}{c}{Bbox (cm)} \\
  No of faces & &  X axis & Y axis &	Z axis	 \\ \hline
  168 & 1.299 & 3.00 &	2.92 &	2.88 \\ \hline
  1520 & 1.399 & 3.00 &	3.00 &	3.00 \\ \hline
  14640 & 1.412 & 3.00 & 3.00 &	3.00 \\ \hline
  148224 & 1.414 & 3.00	& 3.00 &	3.00 \\ \hline
\end{tabular} 
\caption{Geometrical measurements of the sphere primitive.}
\label{tab:spheregeo}
\end{table}

\begin{figure}
\begin{center}
		\captionsetup{justification=centering}
		\includegraphics[width=0.6\textwidth]{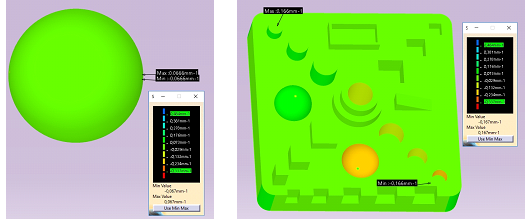}
\end{center}
\vspace{-0.1in}
\caption{ Surface curvature analysis of the initial CAD models (a) sphere  and (b) B3 benchmark.}
\label{fig:curvatures}
\end{figure}

\begin{figure}[h!]
	\begin{center}
		\captionsetup{justification=centering}
		\includegraphics[width=0.9\textwidth]{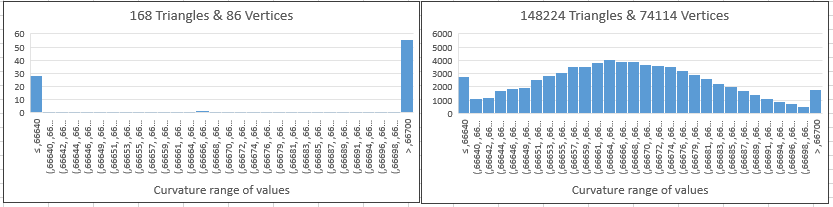}
	\end{center}
	\vspace{-0.1in}
	\caption{Histograms presenting curvature data distribution of a sphere for different triangular meshes.}
	\label{fig:spherehistos}
\end{figure}

\begin{figure} [h]
	\begin{center}
		\captionsetup{justification=centering}
		\includegraphics[width=0.9\textwidth]{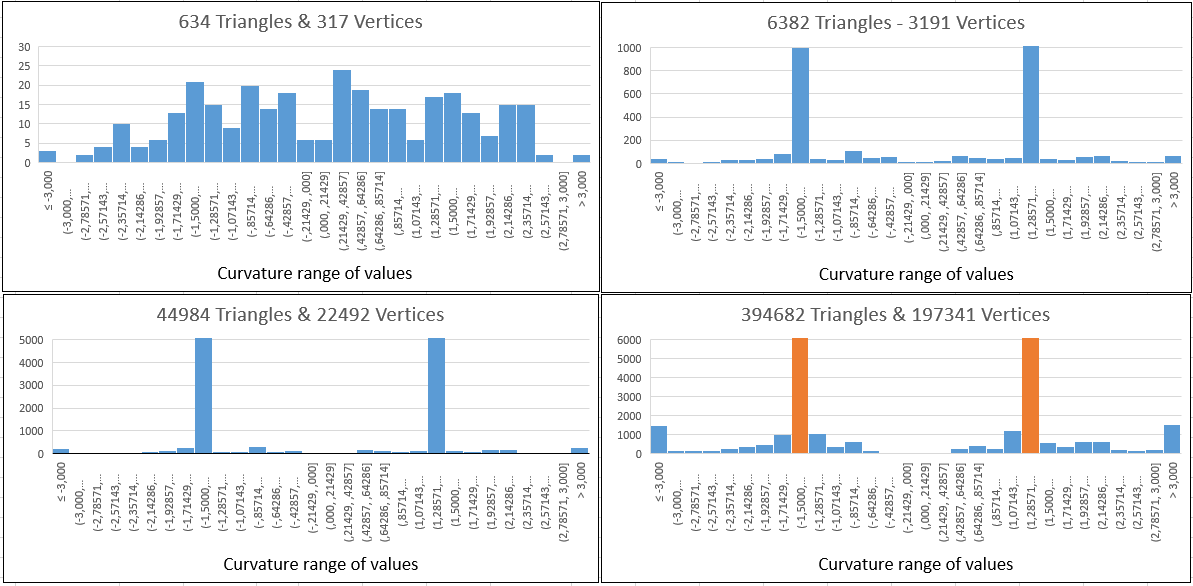}
	\end{center}
	\vspace{-0.1in}
	
	\caption{Histograms presenting curvature data distribution of B3 benchmark for different triangular meshes.}
	\label{fig:B3histos}
\end{figure}

Figure \ref{fig:curvatures} shows the Mean Curvature Analysis performed on the initial CAD models which is correlated with the data range of the different curvatures for each meshed model presented in histograms of Figures \ref{fig:spherehistos} and \ref{fig:B3histos}. More specifically, Figure \ref{fig:spherehistos} displays the curvature data distribution of the meshed primitive sphere for a lower resolution mesh of 168 faces and a higher resolution mesh of 148224 faces. The graph data changes depending on the curvature of each triangle and in the last one it is almost uniform. Finally, the curvature data distribution of the meshed B3 benchmark is displayed in Figure \ref{fig:B3histos}, with the lowest quality corresponding to 634 faces and the highest to 394682 faces. The final distribution is bimodal, with the two peaks representing the two half spheres of the model. As the resolution of the meshed model becomes higher, the deviation with the initial CAD model becomes lower, leading to better results in the final printed model. 

\begin{table}
\centering
\begin{tabular}{c|c|ccc} \hline
  \multicolumn{5}{c}{B3 $V\colon 8,0$ cm \& $H\colon 2,5$ cm \& $D\colon 8,0$ cm- Volume$\colon 9,534*10^{-5}$ $m^3$}  \\ \hline
  Triangular Mesh & Volume ($10^{-5}$ $m^3$) &  \multicolumn{3}{c}{Bbox (cm)} \\ 
  No of faces & 	&  X axis & Y axis & Z axis	 \\ \hline
  634  &  9.511   &  8.00   &  2.50	 & 8.00 \\ \hline
  6382  &  9.534  &  8.00  &  2.50  &  8.00 \\ \hline
  44984  &  9.534  &  8.00 &  2.50  &  8.00 \\ \hline
  394682  &  9.534  &  8.00	&  2.50 &	8.00 \\ \hline
\end{tabular} 
\caption{Geometrical measurements of the  B3 benchmark.}
\label{tab:B3geo}
\end{table}

\begin{figure}[h]
\centering
\begin{subfigure}{.4\textwidth}
  \centering
  \includegraphics[height=.2\textheight]{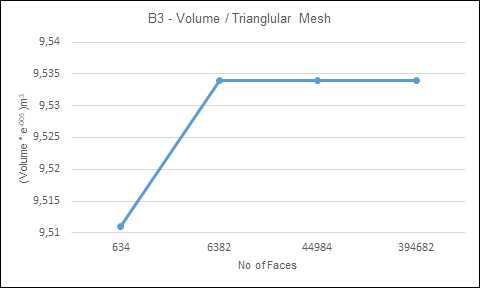}
  \caption{Volume to Triangular mesh diagram}
\end{subfigure}
\begin{subfigure}{.4\textwidth}
  \centering
  \includegraphics[height=.2\textheight]{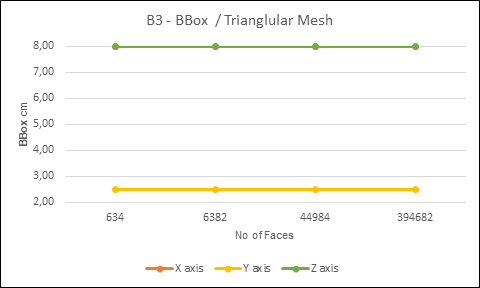}
  \caption{Bbox to triangular mesh diagram}
\end{subfigure}
\caption{Diagrams presenting the effect of triangular mesh over (a) B3 volume and (b) B3 bbox.}
\label{fig:B3graphs}
\end{figure}

\subsection{Design Rules}
Another factor that should be considered when discussing printability is the design process. Designing and printing a CAD model are two separate processes which may result in different outcomes in terms of functionality and geometry. The quality and robustness of a 3D printed model are directly correlated to the AM technology used, the materials and parameters selected for printing and the design rules applied during the initial Computer $-$aided design (CAD) phase. There are design rules that conform with AM technologies and design guidelines that should be considered when designing a model for manufacturing with a specific 3D printing technology\cite{3DHubsDesign}. For example, if a model has an overhang feature of over 45$^{\circ}$ and is to be manufactured using FDM technology, then supports must be added to the structure to ensure printing feasibility. However, when the same model is printed on a powder-bed AM machine support structures are not necessary since support is provided due to the powder itself.

In this work we examine design characteristics and the corresponding design rules that must be enforced for a CAD model to ensure its printability using a specific AM technology. The final printed model depends on the impact of each rule on the final outcome. A print failure may occur because of structural problems (e.g. a collapsed wall), dimensional accuracy deviations (e.g. holes with a smaller diameter) or functionality and assembly issues (e.g. parts that must fit together, e.g. a screw). Printing problems may also arise in models with high level of detail on a small surface part (see e.g. filigree jewelry~\cite{Stamati2011}) due to AM technology limitations. Another type of "poor printability" is encountered when many support structures must be included to facilitate printing. These supports, when removed, may lead to rough surfaces or damages to the core model. 

\begin{figure}
\centering
	\begin{centering}
		\captionsetup{justification=centering,width=0.89\textwidth}
		\includegraphics[width=0.89\textwidth]{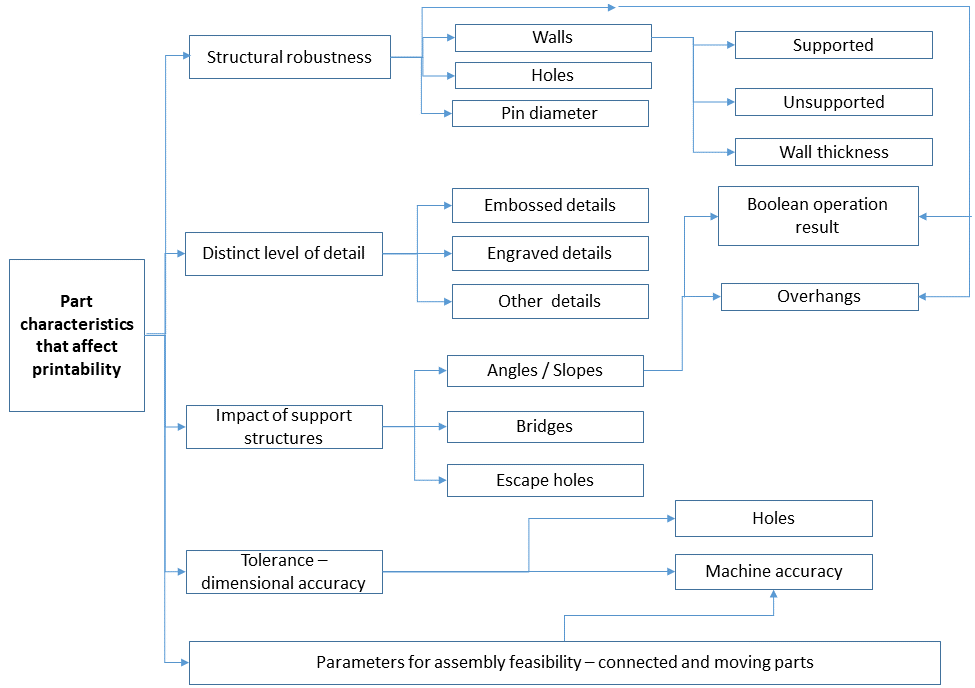}
	\end{centering}
	\vspace{-0.1in}
	\caption{Part design characteristics.}
	\label{rules}
\end{figure}

To this end, we propose a suite of part characteristics based on best design practices depicted in Figure \ref{rules}. More specifically, there are design rules that, when applied, ensure the structural robustness of the model. Part characteristics that fall into this category are walls (supported, unsupported with a minimum wall thickness), holes (minimum diameter), pins (min diameter), Boolean operation parameters (e.g. overlapping components united to form a larger one) and overhangs. These part characteristics must conform with size limitations that are derived from the manufacturing accuracy of various AM technologies to ensure printability. 
Design parameters are also defined for achieving distinct level of detail for parts that are embossed or engraved, i.e. text reliefs on a manufactured part. There are design guidelines that pertain to support construction, which mainly concerns a subset of AM technologies. Supports should be added in the case of overhangs and bridges (for some AM machines) for the model to be printable. Support construction on its own adds complexity to the printing process in terms of cost and post processing labor. 
Lastly, there are design rules that refer to tolerance and dimensional accuracy issues and practices that ensure the functionality of connected and/or moving parts. 
In this work we shall provide examples of characterizing 3D printability by translating such rules and practices with respect to the FDM, 3DP and Polyjet 3D printing technologies. Our framework can be used to characterize printability in other AM technologies as well by tuning model parameters. 

\cleardoublepage
\section{A CHARACTERIZATION OF 3D PRINTABILITY}
\label{sec:printability}
In this work we define a measure that characterizes the printability $P$ of a model $M$ on a 3D printing technology $T$. This printability score is expressed by a number on a scale of 0 to 100, where 0 is a model that will result in a print failure on a specific AM technology and 100 corresponds to a model that is structurally robust and flawless when printed using a specific AM technology. The printability score is defined by two factors: the global probability function and the part characteristic probability function.

The global probability function expresses the probability of printing problems due to the printing technology characteristics of the AM machine to be employed and the model mesh complexity. If $P_{G}(C_{M},T,A)$ is the global probability function for an unsuccessful print on a specific technology, based on the mesh complexity $C_{M}$ of the model, the technology $T$ used and the application $A$ domain of the model, then $(1-P_{G}(C_{M},T,A))$ is the corresponding probability function for a successful print.

A meshed model {\textit{M}} can also be described by a set of part characteristics $i$, that affect the printability of the model. If $P_{F}(i,D,T,A)$ is the probability of a part characteristic $i$ with a set of characteristic parameters $D(i)$ to exhibit a flaw regarding an application $A$ that will affect the entire printed model for a specific printing technology $T$, then $(1-P_{F}(i,D,T,A))$ is the probability of the part characteristic to lead to a successful overall printing result. The overall probability of a model $M$ with $n$ number of part characteristics to be successfully printed on technology $T$ is:
\begin{equation}
\label{eq:overall}    
   P(M,T)= (1-P_{G}(C_{M},T,A)) *\prod_{i=1}^{n} (1-P_{F}(i,D,T,A))
\end{equation}

Then the printability measure (score) of $M$ on $T$ is:
\begin{equation}
\label{eq:printability}  
   PS(M,T)=100*P(M,T)
\end{equation}


\subsection{Global Probability Function}
The global probability function is related to the characteristics of the technology employed for printing. Objectively, each technology has advantages and disadvantages~\cite{ntousia2019}. A selection of technology characteristics that were deemed as very significant was made and is summarized in Table ~\ref{tab:printerCharacteristics}. An initial defect score was assigned to each characteristic $x$ based on technical specifications of each technology $T$ and experimental technology assessment presented in previous research \cite{Ramya2016}. This score was subsequently transformed to  $DS_{T_{Perfect}}(x)$ that expresses the  probability of a characteristic $x$ to cause a significant problem in the printing result for the highest mesh resolution.
A defect score of "***" means that the characteristic presents a low probability to have a negative effect on the printing process and therefore, a small probability value (e.g. 0.01) will be assigned to this characteristic. On the other hand, a defect score of "*" means that this characteristic exhibits a high probability to have a negative effect on the process and, as such, a higher probability function value (e.g. 0.5) is assigned. The probability values $DS_{T_{Perfect}}(x)$ are determined using the characteristics of technology $T$.
In the scope of this work, the defect scores are correlated with the following defect probability values: "***"=0.01 (1 \% probability), "**"=0.03 (3 \% probability), and "*"=0.05 (5 \% probability).

The aforementioned defect score probability values for the 3D printing characteristics of the technology provide a characterization of the quality ranking of the technology. The values chosen for the scoring schema can be altered depending on the requirements and restrictions desired for determining printability in terms of a specific application. For an application dependent on assembly and connected parts functionality  the defect score for accuracy is the one determined by the technology. In contrast, for artistic application the accuracy defect score can be alleviated by multiplying by a number $k$ as explained below.  

\begin{table}[h!]
\centering
\begin{tabular}{>{\raggedright\arraybackslash}m{4.5cm}|ccc}  \hline
Characteristic $/$ Technology & FDM & Binder Jetting & Material Jetting\\ \hline
Accuracy & ** & ** & *** \\
Surface Texture & * & ** & *** \\ 
Various Abnormalities (Warping, shrinkage etc.) & * & ** & ***\\
Support Construction & ** & *** & ** \\ \hline
\end{tabular}
\caption{Selection and scoring of printing technologies: ***=best quality, **=average quality, *=lower quality.} 
\label{tab:printerCharacteristics}
\end{table}

We define the defect score probability function $DS_{T}(x)$ of a technology printing characteristic $x$ on technology $T$ as (Equation \ref{eq:defectscorex}):
\begin{equation}
\label{eq:defectscorex}
DS_{T}(x)=1-(1-DS_{T_{Perfect}}(x))*QS_{C_{M}}
\end{equation} 
where $QS_{C_{M}}$ is a quality measure assigned to the model based on the ratio of the mesh surface area $Area(M)$ to the surface area $Area(O)$ of the original CAD model (Equation ~\ref{eq:areaRatio}). This ratio is based on the mesh complexity of the model and expresses the error introduced by mesh triangulation. Models with low mesh complexity produce parts of lower quality whereas meshes of higher complexity lead to products of a much better quality and robustness. For the calculation of the defect score probability function for the support construction characteristic $DS_{T}(Supp)$, we consider $QS_{C_{M}}=1$ because mesh resolution is not correlated to support structure usage.
\begin{equation}
\label{eq:areaRatio}  
   QS_{C_{M}}= Area(M)/Area(O)
\end{equation}

The global probability function of a model $M$ for an application $A$ on a printing technology $T$ is evaluated by Equation~\ref{eq:global}:
\begin{equation}
\label{eq:global}
\begin{split}
   P_{G}(C_{M},T,A) &= 1-\prod_{x \in S} (1-DS_T(x)*k(x,A)
   \end{split}
\end{equation}

 where $S=\{Accuracy,Surface Texture, Various Abnormalities, Support Construction\}$ is the set of global technology characteristics and $k(x,A) \in [0,1]$ is a factor that determines the sensitivity of application $A$ to characteristic $x$, where a value of $k(x,A)=0$ means that application $A$ is not affected by the characteristic $x$, whereas $k(x,A)=1$ means that application $A$ is fully affected by chacteristic $x$.

 Example: Given the CAD model of a sphere (diameter 30mm) and four different mesh representations (STL files of different resolutions), the global probability function for each model is computed on the three AM technologies, based on the information in Table ~\ref{tab:printerCharacteristics}. The results are depicted in Table ~\ref{tab:globalExampleSphere}. For $k$ = 0.1, the model with the lowest mesh complexity (168 triangles) is presented with a slightly lower probability function than the more dense meshes, especially on FDM technology. However, for a higher value of $k$, (e.g. $k=0.5$) this difference in the mesh complexity is more significant, evidently affecting the global probability function values. For $k=0$, $P_{G}(C_{M},T,A)=1$.
 
\begin{table}[h!]
\centering
\small
\begin{tabular}{l|lll||lll} \hline
  \multicolumn{7}{c}{Global probability function: $1-P_{G}(C_{M},T,A)$ } \\ \hline
 $C_{M}$ & FDM & Binder Jetting & Material Jetting & FDM & Binder Jetting & Material Jetting\\ \hline
 168 & 0.97239246 & 0.978116384 & 0.981823174 & 0.867419752 & 0.89389853 & 0.911550807\\ \hline
1520 & 0.982638713 & 0.988553692 & 0.992496154 &0.915394871 & 0.943702897 & 0.962885818 \\ \hline
14640 & 0.983937057 & 0.989876312 & 0.993848738 & 0.921572972 & 0.950117484 & 0.969498937\\ \hline
148224 & 0.984078112 & 0.990020005 & 0.993995687 & 0.922245513 & 0.950815781 & 0.970218865\\ \hline
\end{tabular} 
\caption{Global probability function values expressing printability for mesh models of a sphere with 168, 1520, 14640 and 148,224 triangles (left) with $k$=0.1 and (right) $k$=0.5.}
\label{tab:globalExampleSphere}
\end{table} 
 
\subsection{Part Characteristic Probability Function}
For each design part characteristic $i$  we determine a part characteristic probability function (PCP function) $P_F$ that depends on the printing technology $T$, the design characteristic $i$ with a set of characteristic parameters $D(i)$  and the application $A$.
A part characteristic $i$ that falls under the categorization depicted in Figure \ref{rules} is susceptible to flaws occurring for each design rule that is violated (since this increases the probability of a print failure). To determine the flaw probability function $P_F$ of a part characteristic we consider the following parameters: (i) The weight $w(T,i) \geq 0$ is a numerical parameter that depends on the technology and the design characteristic, and is the dimension value of the design characteristic $i$ that has probability $50\%$ to exhibit a significant flaw during printing on technology $T$. This parameter can be determined by the thresholds reported by various practitioners or researchers (see e.g. \cite{3DHubsDesign}). A threshold value for a dimension (for example in thin parts) expresses a point were we have a $50\%$ probability of failure. (ii) The significance $0 < s(A,i) \leq 1$ expresses the impact of the corresponding design characteristic $i$ on the printed model regarding application $A$. This is set by the designer for each part characteristic and represents the effect of the violation of a design rule for a part characteristic on the functionality and the robustness of the overall product. For example, the significance of a through hole is small for decorative artifacts and large if it is used for ventilation or assembly of a mechanical part.

The PCP function of a part characteristic that corresponds to thin parts or small holes can be described by Equation \ref{eq:pcpf}:

\begin{equation}
   P_{F}(i,d,T,A) = (1- \frac {1}{1+e^{w(T,i)-d}})*s(A,i)
   \label{eq:pcpf}
\end{equation}

where $i$ is the characteristic under evaluation, $d$ is its dimension (for holes and thin parts $D(i)=\{d\}$), $w(T,i)$ and $s(A,i)$ are the two numerical parameters described above.

PCP functions for other characteristics can be derived using the approach presented in this section. An additional factor that should be used for support characteristics is the ratio of the surface of the model affected by support structures over the entire model surface area.

\subsection{Setting Parameters and Thresholds}
\label{sec:dicsussion}

Most parameters of the printability measure parameters are either derived from experiments conducted by researchers or practitioners or can be determined experimentally by experts. The only set of parameters set by the user-designer is the sensitivity of his/her application to defects triggered by each characteristic. For the entire process the designer needs to determine: (i) how sensitive is the application to global technology characteristics (set $k(x,A)$) and (ii) the effect for each part design characteristic $i$ on the robustness of the entire model for application $A$ (set $s(A,i)$). For example a designer of an artistic object should set the corresponding global sensitivities to the maximum value (i.e. 1) for {\em surface texture} and {\em supports} and the minimum value (i.e. 0) for the rest of the global characteristic. For part characteristics the user designer of an artistic object should set the maximum value to parts that the failure of the corresponding design rule will cause a collapse of a significant part of the object (i.e. thin walls) and can ignore through holes and other decorative elements that have a small impact on the aesthetics of the final object.

After setting these parameters the system will be able to determine automatically the overall $printability$ score. For example a model with a printability score of less than $80\%$ has a high chance of exhibiting structural robustness problems and/or undesired characteristics (i.e. warping, rough surfaces). The strict threshold under which a model will be characterized as 'unprintable' can be defined by the needs of the end-user and the cost of printing. For example a printability score of $75\%$ means that only one out of four prints of a specific design will fail due to any characteristic or design rule, which is perfectly fine if we have a technology with medium cost of consumables. Finally, for making extremely complicated prototypes even a threshold of $30\%$ could be acceptable since this means that one out of three attempts to print the specific model will succeed.


\section{VALIDATION OF THE PRINTABILITY MEASURE THROUGH TEST CASES}
\label{sec:validation}
The evaluation of our proposed scoring method was performed using three AM machines using different technologies: FDM, 3DP and Polyjet. The FDM 3D printer used in this study is the Ultimaker 3 Extended. Its dimensional accuracy is $0.2 - 0.02\:mm$ for a $0.4\:mm$ nozzle and it was used with Polylactic Acid (PLA) filament as a feedstock material. The 3DP, Binder Jetting technology based AM machine used in our research is the ZCorp 450 3D printer and its dimensional accuracy is $+/- 0.102\:mm$. We have also tested our models on the Stratasys Connex3 Objet 260, representing Polyjet technology, and having a dimensional accuracy of Up to $200\: \mu m \: (0.2\: mm)$.

\begin{figure}[h]
	\begin{center}
		\captionsetup{justification=centering}
		\includegraphics[width=0.6\textwidth]{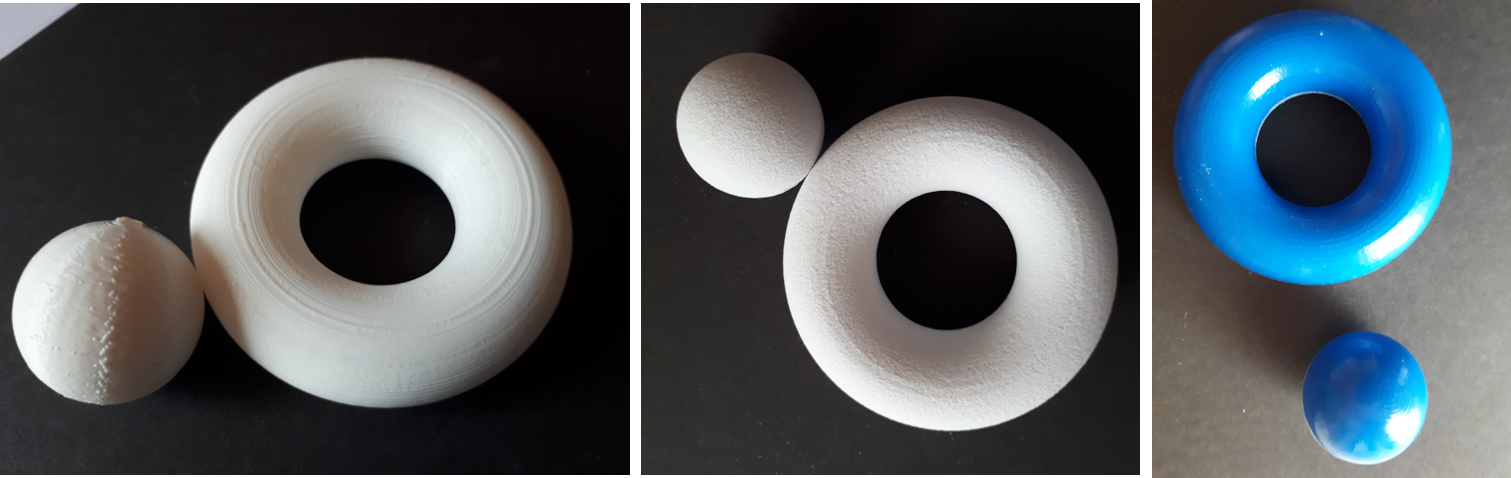}
	\end{center}
	\vspace{-0.1in}
	\caption{Examples of printed primitives on FDM, 3DP and Polyjet technologies.}
	\label{fig:primitives}
\end{figure}

The test models used for evaluation were geometric primitives and the three benchmark models (see Section \ref{sec:characteristics}). Each geometric primitive was printed 5 times on each AM machine and each benchmark was printed 3 times. Sample photos of the tori and spheres printed on the three AM machines are displayed in Figure ~\ref{fig:primitives}. The printability score for each model on each technology was calculated before printing (Table ~\ref{tab:testcases}), having $k=0.1$ for the global probability function. We evaluated the printability scores tuning the parameters such that a high sensitivity was set for holes and thin parts. The geometric primitives do not contain any part characteristics that may lead to a print failure, therefore their part characteristic probability functions are equal to $0$. Given this, their printability scores are derived only from their global probability functions. Based on the printability scores, the sphere on FDM technology has a higher probability to display printing errors. Also, the cylinder, rectangular parallelepiped and torus have a printability score over 99\% on Polyjet technology and 3DP. As for the printability scores of the benchmark models, the part characteristic probability functions are evaluated for thin walls, pins and holes.

\begin{table}[h!]
\centering
\begin{tabular}{l|ccc} \hline
  & \multicolumn{3}{c}{Printability Score ($k=0.1$)} \\ \hline
 Model & FDM & Binder Jetting & Material Jetting \\ \hline
Sphere & 98.379\% & 98.973\% & 99.370\% \\ \hline
Cylinder & 98.406\% & 99.000\% & 99.397\% \\ \hline
Torus & 98.409\% & 99.004\% & 99.401\% \\ \hline
Rect. Parallelep. & 98.406\% & 99.001\% & 99.398\% \\ \hline
B1 & 22.110\% & 12.100\% & 28.425\% \\ \hline
B2 & 28.679\% & 17.592\% & 39.421\% \\ \hline
B3 & 86.998\% & 74.239\% & 86.025\% \\ \hline
\end{tabular} 
\caption{Printability score of printed models}
\label{tab:testcases}
\end{table} 
 
After printing and post processing, we evaluated the fabricated parts from different perspectives: dimensional accuracy, structural robustness and surface quality. 
 
 \begin{figure}[h]
	\begin{center}
		\captionsetup{justification=centering}
		\includegraphics[width=0.9\textwidth]{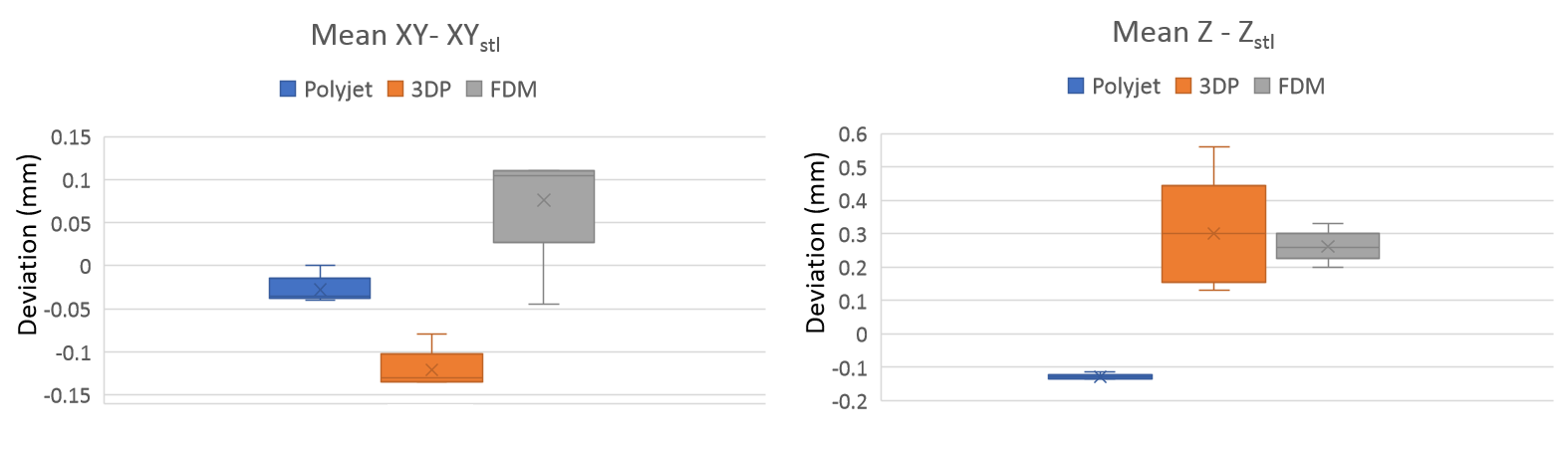}
	\end{center}
	\vspace{-0.1in}
	\caption{Display of XY and Z deviations of printed sphere from original sphere mesh.}
	\label{fig:sphere-deviations}
\end{figure}
 
 To evaluate the dimensional accuracy of all the fabricated parts and the overall conformity to the original models, we performed measurements with a digital caliper, obtaining at least 10 measurements for each dimension. These measurements were then processed to calculate the mean value for each dimension. With these values we calculated the deviation in reference to the dimensions of the original mesh model that was used for fabrication. In general, the parts fabricated using Polyjet technology were more accurate and displayed repeatability and consistency, in contrast to the other technologies. For example, when measuring the sphere, measurements we obtained from both the XY axis and the Z axis. The results for the sphere are displayed in Figure \ref{fig:sphere-deviations}. The parts printed with Polyjet technology where the most accurate in both XY and Z. The 3DP parts were smaller than the original on the XY axis, which could be explained by the lower resolution of the machine on XY, whereas on the Z axis they were larger. The parts printed on FDM were generally larger in both directions. Also, a characteristic of the FDM parts was that they displayed warping. For example, when evaluating the diameter of the cylinder, a distinct differences were recorded throughout its length. 
 
 For the geometric primitives we calculated the ratio $l$ of the volume of the fabricated part $A$, $V_A$, to the volume of the initial mesh model $M$,$V_M$, to measure the actual accuracy of the AM machine (Equation \ref{eq:vol}). In general, the objects that were printed on Binder Jetting technology had the smallest deviation in volume size from the original model (Table ~\ref{tab:spherevolume}).
 
\begin{equation}
l = \frac{V_M}{V_A}
\label{eq:vol}
\end{equation}

\begin{table}[h!]
\centering
\begin{tabular}{l|ccc} \hline
  & \multicolumn{3}{c}{Ratio $l=V_M/V_A$} \\ \hline
 Model & Polyjet & 3DP & FDM \\ \hline
 Sphere-1 &  0.993682194  &  1.010465886 &  1.021082769  \\ \hline
 Sphere-2 & 0.993433109	& 1.001674907 &	1.017535587	 \\ \hline
 Sphere-3 &  0.993433109 & 1.026164459 & 1.012230231 \\ \hline
 Sphere-4 & 0.993682194 & 1.004181391 &	1.023875618   \\ \hline
 Sphere-5 & 0.996175332	& 1.012230231 &	1.020068449  \\ \hline
\end{tabular} 
\caption{Volume ratios of printed sphere models.}
\label{tab:spherevolume}
\end{table}

\begin{figure}[h]
	\begin{center}
		\captionsetup{justification=centering}
		\includegraphics[width=0.9\textwidth]{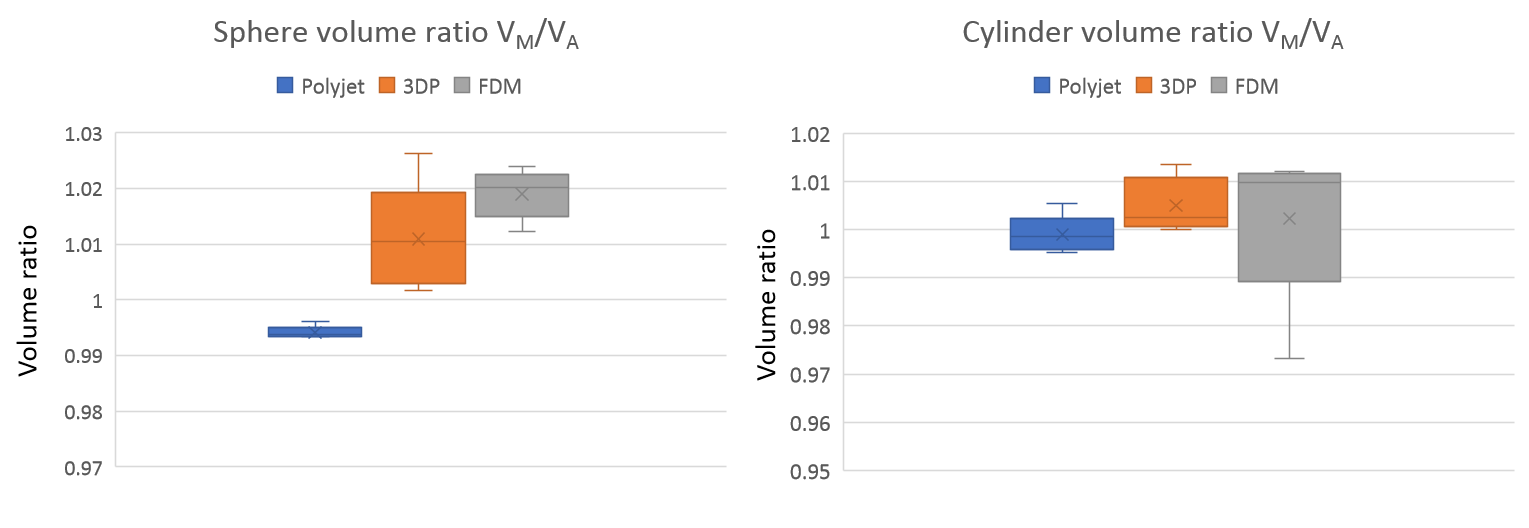}
	\end{center}
	\vspace{-0.1in}
	\caption{Comparison of the volume ratios for the parts fabricated on the three technologies.}
	\label{fig:volumeRatios}
\end{figure}

\begin{figure} [h]
\centering
\begin{subfigure}{.76\textwidth}
  \centering
  \includegraphics[width=.76\linewidth]{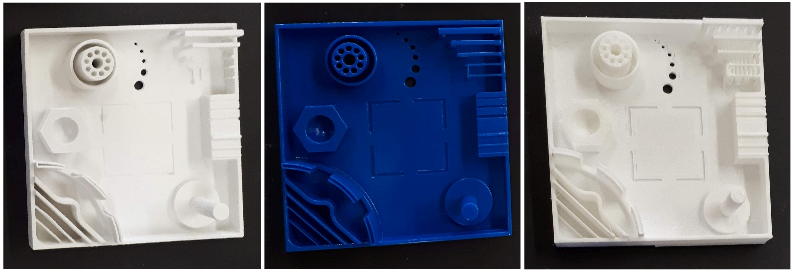}  
  \caption{3DP (left), Polyjet (middle) and FDM (right) prints of the B1 benchmark}
\end{subfigure}
\begin{subfigure}{.76\textwidth}
  \centering
  \includegraphics[width=.76\linewidth]{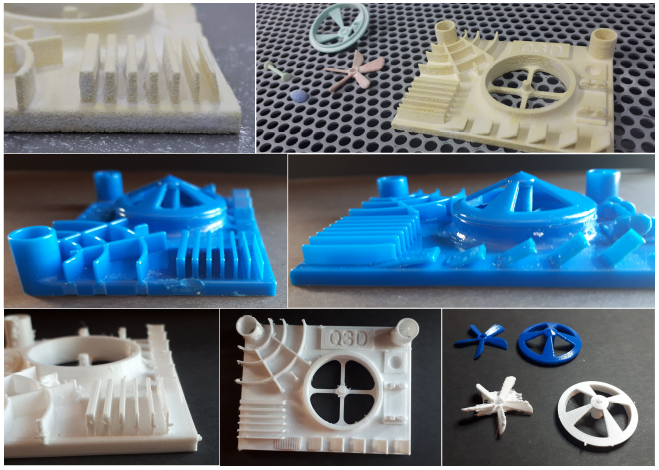}
  \caption{3DP (top), Polyjet (middle) and FDM (bottom) prints of the B2 benchmark}
\end{subfigure}
\begin{subfigure}{.76\textwidth}
  \centering
  \includegraphics[width=.76\linewidth]{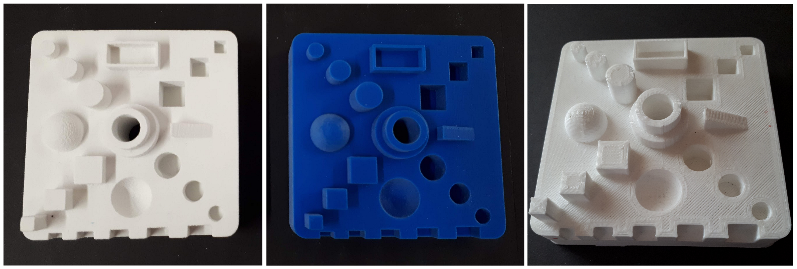} 
  \caption{3DP (left), Polyjet (middle) and FDM (right) prints of the B3 benchmark}
\end{subfigure} 
\caption{B1, B2 and B3 benchmarks printed on the three technologies.}
 \label{fig:Bench-prints}
\end{figure}

 In reference to the structural robustness of the fabricated parts, no issues were found in the geometric primitives. In benchmark B1, there were structural problems in some models. In the FDM prints, the thinnest bridges could not be cleaned of the supports; there was a structural collapse when this was attempted. The same problem arose in 2 of the 3 Polyjet parts. In the 3DP model, the thin bridges were printed, however they broke in post processing. In all models the holes were printed, however in the FDM parts some holes were not circular. As for benchmark B2, the overhangs with the 2 smallest angles were not printed well on FDM, and in the 3DP models there was a print failure whilst cleaning. The thinnest wall was successfully printed on Polyjet, on FDM there was a drop in the material, whereas on 3DP it broke in one of the three models, while the other two presented warping (Figure \ref{fig:Bench-prints}(b)). The thin pin was printed on all technologies, however in the 3DP model it collapsed during cleaning. Also, the propeller of B2, which is a thin structure, was successfully printed on all technologies, however cracked during support removal of the FDM part. Benchmark B3 was successfully printed on all AM machines with no structural problems.

As for the surface texture and aesthetic result, there is distinct difference in quality. The FDM parts are rough, with more surface anomalies, hairs, uneven surfaces from material deposition. 3DP parts are slightly porous but with a better level of detail and no anomalies due to lack of supports. The Polyjet prints have a very smooth surface with good level of detail.

Given the above observations, we reviewed the printability scores of the models. In reference to the geometric primitives (Figure ~\ref{fig:primitives}), the printing scores conform with the result: all four primitives are printable, without features that can cause a structural failure. The difference in printability scores on all technologies for a model, e.g. the sphere, is justified, through the measurements but also visually, as shown in Figure ~\ref{fig:primitives}. The sphere printed on the FDM machine has surface abnormalities that are not present in the other parts. 

As for the benchmark models, B1 had a printability score that was low, due to the presence of many thin parts whose dimensions were at or  below the limits of the AM technologies and this is justified by the fact that there was breakage of thin parts on all AM machines. Model B2 had a higher printability score than B1, but still relatively low, and this proved correct since there were print failures of thin walls. The printability score of B3 was much better, and therefore the model was robustly printed on all technologies. All 3 printed benchmarks are shown in Figure  \ref{fig:Bench-prints}.

\section{CONCLUSIONS}

In this paper we have proposed a novel approach to characterizing the efficacy of manufacturing a designed CAD model on an AM machine of a certain technology, based on its model complexity and part characteristics. These elements are mapped to parameters and functions, that depend also on the printing technology to be employed, that make up a linear formula that corresponds to a {\em printability} score. This measure, which is evaluated using worst case printing scenarios, can be used either to determine which 3D technology is more suitable for manufacturing a specific model or can be used as a guide to redesigning the model so that it is more suitable for an intended specific technology. 

As future work we intend to evaluate more part characteristics and their impact on printability. We will evaluate the volume ratios of the benchmark models with methods of photogrammetry and laser scanning to further validate our approach. Also our proposed printing score system can be adapted to include other AM printing technologies and other design intents. 

\section*{ACKNOWLEDGEMENTS}
This research has been co-financed by the European Union and Greek national funds through the Operational Program Competitiveness, Entrepreneurship and Innovation, under the call RESEARCH -CREATE -INNOVATE (project code:T1EDK-- 04928)

\referenceSection
\bibliographystyle{CADA}
\bibliography{CADandA_Paper_Template}

\bigskip
\end{document}